\documentclass[%
 reprint,
superscriptaddress,
nofootinbib,
 amsmath,amssymb,
pra,
]{revtex4-2}

\usepackage{graphicx}% Include figure files
\usepackage{dcolumn}% Align table columns on decimal point
\usepackage{multirow}
\usepackage{upgreek}
\usepackage[uncertainty-mode=separate]{siunitx}
\usepackage{booktabs}
\usepackage{ragged2e}
\usepackage{bm}% bold math
\usepackage[hidelinks=true]{hyperref}% add hypertext capabilities
\usepackage[capitalise, nameinlink]{cleveref}

\crefname{section}{Sec.}{Sec.}
\Crefname{section}{Sec.}{Sec.}
\newcommand{\um}{\unit{\upmu\meter}}
\newcommand{\us}{\unit{\upmu\second}}

\begin{document}

\title{Demonstrating Single Photon Counting with \\Kinetic Inductance Detectors from 3.8 to 25 $\upmu$m}

\author{Wilbert G. Ras-Vinke}%
    \email[Contact author at ]{w.ras@sron.nl}
    \affiliation{Space Research Organisation Netherlands (SRON), Niels Bohrweg 4, Leiden 2333 CA, Netherlands}
    \affiliation{Department of Microelectronics, Delft University of Technology, Mekelweg 4, Delft 2628 CD, Netherlands}
\author{Kevin Kouwenhoven}%
    \affiliation{Space Research Organisation Netherlands (SRON), Niels Bohrweg 4, Leiden 2333 CA, Netherlands}
    \affiliation{Department of Microelectronics, Delft University of Technology, Mekelweg 4, Delft 2628 CD, Netherlands}
\author{Jochem J.A. Baselmans}%
    \affiliation{Space Research Organisation Netherlands (SRON), Niels Bohrweg 4, Leiden 2333 CA, Netherlands}
    \affiliation{Department of Microelectronics, Delft University of Technology, Mekelweg 4, Delft 2628 CD, Netherlands}
    \affiliation{Physikalisches Institut, Universität zu Köln, Zülpicher Straße 77, 50937 Cologne, Germany }
\author{Kenichi Karatsu}%
    \affiliation{Space Research Organisation Netherlands (SRON), Niels Bohrweg 4, Leiden 2333 CA, Netherlands}
\author{David J. Thoen}%
    \affiliation{Space Research Organisation Netherlands (SRON), Niels Bohrweg 4, Leiden 2333 CA, Netherlands}
    \affiliation{Department of Microelectronics, Delft University of Technology, Mekelweg 4, Delft 2628 CD, Netherlands}
\author{Vignesh Murugesan}%
    \affiliation{Space Research Organisation Netherlands (SRON), Niels Bohrweg 4, Leiden 2333 CA, Netherlands}
\author{Pieter J. de Visser}%
    \affiliation{Space Research Organisation Netherlands (SRON), Niels Bohrweg 4, Leiden 2333 CA, Netherlands}
    \affiliation{Department of Microelectronics, Delft University of Technology, Mekelweg 4, Delft 2628 CD, Netherlands}
    
\date{\today}

\begin{abstract}
One of the primary objectives of modern astronomy is the atmospheric characterization of Earth-like exo\-planets at visible and infrared wavelengths. Achieving this goal requires extremely sensitive detectors capable of measuring the exoplanet’s faint signal at the single-photon level while maintaining near-zero dark count rates. In the infrared, however, conventional semiconducting detector technologies struggle to meet these stringent requirements. In this work we demonstrate single-photon counting with superconducting, Microwave Kinetic Inductance Detectors at the wavelengths \qtylist[list-units=single]{3.8;8.5;18.5;25}{\um} and measure resolving powers $\left(E/\delta E\right)$ of \numlist{9.9;5.9;3.2;3.3}, respectively, with corresponding dark count rates of \qtylist[list-units=single]{4;8;34;48}{mHz}. Our membrane-based devices reach phonon-loss limited performance at \qty{3.8}{\um}, more than doubling the performance attainable with comparable solid-substrate devices. These results showcase the detector technology in the mid-infrared and the intricate measurement setup required for these sensitive detectors. We discuss how the detector design and measurement setup can be further optimized to increase the detector performance in the mid-infrared.

\end{abstract}

\maketitle
\section{Introduction}
\label{sec:introduction}
\begin{figure*}[ht!]
    \includegraphics[width=\linewidth]{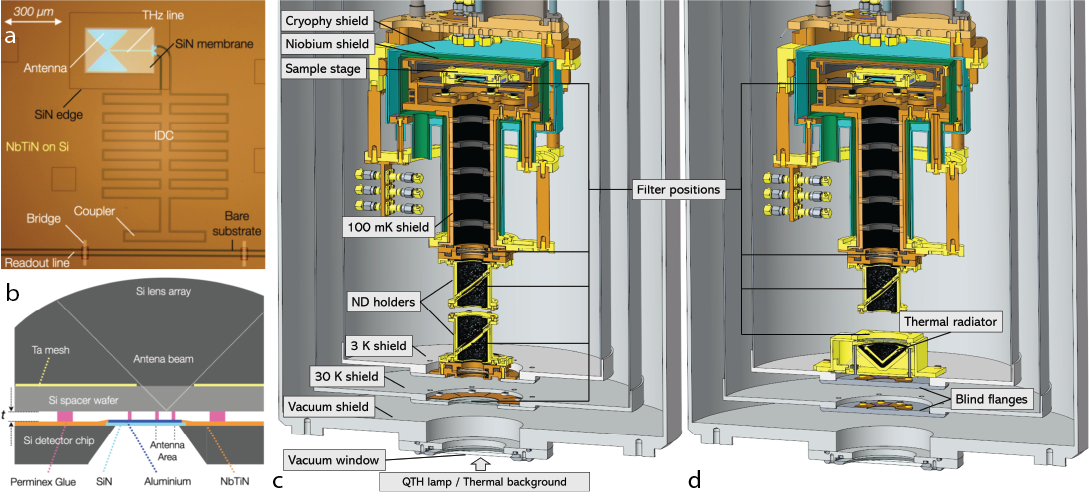}
    \caption{MKID architecture and cryogenic setup. (a) Geometry of a single detector. The microwave resonator is made of a NbTiN IDC in parallel with aluminium CPW (THz line) with a highly inductive central line. The THz line is at the focus of a Si lens which is also the feed point of a leaky-slot antenna. The THz line is suspended on a SiN membrane to decrease phonon loss to the substrate. The detector design is originally optimized for the far-IR at \qty{200}{\um}. (b) Cross-sectional view of the detector assembly. The figures in panels a and b are reproduced from Ref. \cite{baselmans2022ultra}. (c) Cross section of the DR as configured for the experiments at \qtylist[list-units=single]{3.8;8.5}{\um} and (d) at \qty{18.5}{\um}. The detectors are mounted at the sample stage. Indicated are the temperature shields as well as the Cryophy and Niobium magnetic shields. The key difference between the configurations is the source of radiation; (c) an external QTH lamp and (d) a cryogenic radiator at \qty{3}{K}. Optical filters can be mounted at the indicated positions.}
\label{fig:mkid}
\end{figure*}

Key to our understanding of life in the universe is the atmospheric characterization of Earth-like exoplanets to learn about their diversity, habitability and ultimately, whether they harbor life \cite{astro2020}. New ground- and space based observatories are currently in development that aim to do direct exoplanet imaging and spectroscopy in the visible to near-infrared (\qtyrange{0.4}{2.5}{\um})\cite{bailey2023nancy, luvoir2019luvoir, gaudi2020habitable, simard2013thirty, codona2004exoplanet, kasper2021pcs}, as well as in the mid-infrared (mid-IR, \qtyrange{2.5}{30}{\um}) \cite{metis2021metis, quanz2022large,dannert2022large}. In this work we focus on the mid-IR wavelength range to address a gap in detector technology for these wavelengths.

The mid-IR is a scientifically valuable wavelength regime for a number of reasons. First, the mid-IR contains important absorption lines that are complementary to lines in the visible and near-IR and help constrain the atmospheric models. Second, the mid-IR offers a much more favorable (order \num{1000}) star-planet contrast than the visible regime, alleviating the requirements on the optical system. Lastly, the planet's brightness in the mid-IR puts better constraints on it's temperature and radius.\cite{kaltenegger2017characterize} However, the expected signal from such an exoplanet is so weak that it will require single-photon sensitive detectors with near-zero dark counts \cite{konrad2022large}. Conventional detectors based on HgCdTe, InAs/GaSb or Si:As technologies are unable to meet the required sensitivity in the mid-IR \cite{roellig2020mid, taylor2023low, hampel2025tungsten}. There are semiconductor-based detectors that reach single-photon sensitivity, but these do not cover the full mid-IR wavelength range \cite{lau2023superconducting, dello2022advances, huber2024apd, anderson2022apd}. 

Meanwhile, important progress has been made with superconducting single-photon detectors for the mid-IR. Superconducting Nanowire Single-Photon Detectors \cite{gol2001picosecond} have demonstrated single-photon counting from the visible up to \qty{29}{\um} with dark counts below \qty{100}{mHz} \cite{taylor2023low} and saturated internal detection efficiencies \cite{hampel2025tungsten}. Microwave Kinetic Inductance Detectors (MKIDs) \cite{day2003broadband}, on the other hand, have now demonstrated single-photon counting at \qty{25}{\um} with an optical efficiency of \qty{46}{\%} and dark count rate of \qty{5}{mHz} \cite{day202425}. MKIDs are superconducting detectors with a lot of heritage in astronomical application---both in the sub-mm to far-IR (\qtyrange{2}{0.03}{mm}) \cite{calvo2016nika2, endo2019first, reyes2025amkid} and in the visible to near-IR \cite{walter2020mkid}---due to their relative ease of multiplexing and fabrication. MKIDs for the sub-mm/far-IR are typically operated in a power-integrating mode with a lens-antenna structure to efficiently couple the light to the sensitive part of the detector, usually made of aluminium. However, the low resistivity of aluminium and limitations in fabrication technology make it progressively harder to achieve a high optical efficiency at shorter wavelengths. Creating periodic meandering absorbers from sub-micron wide thin aluminium can create a wideband absorber that is more robust to misalignment and assembly issues \cite{dabironezare2025lens}. Such lens-absorber MKIDs have demonstrated single-photon detection at \qty{25}{\um} \cite{day202425} and highly efficient power detectors at \qty{43}{\um} \cite{dabironezare2025lens}. However, aluminium cannot be used to make an efficient detector in the visible/near-IR and is often replaced with higher resistivity superconductors, like beta-phase Tantalum, Hafnium or Titanium  \cite{mazin2022superconducting, kouwenhoven2023resolving}. Still, the highest energy resolving powers in the visible/near-IR have been achieved by suspending aluminium on a thin membrane \cite{devisser2021phonon}. The membrane reduces the loss of energetic phonons to the substrate which can increase the MKID's energy resolving power---if it is phonon-loss limited---by a factor \num{2.4}. Currently, energetic phonon-loss to the substrate is limiting the energy resolving power at \qty{25}{\um} to \num{2.9} \cite{day202425}. Phonon-trapping membranes will not give MKIDs the energy resolving powers required for spectroscopy, but the enhanced energy resolving powers will effectively reduce the detector’s dark count rate by improving its ability to discriminate between in-band and out-of-band photons \cite{swimmer2023characterizing}.

In this work we demonstrate single-photon counting with MKIDs on phonon-trapping membranes at \qtylist[list-units=single]{3.8;8.5;18.5;25}{\um}. We measure the energy resolving power at every wavelength and study its different contributions. We also describe and characterize the versatile and sensitive measurement setup that is required to facilitate single-photon counting across this wide wavelength range. The detector, measurement setup and pulse analysis are described in \cref{sec:methods}. In \cref{sec:results} we present and discuss the results on the energy resolving powers and dark count rates. We summarize the outcomes and discuss future work in \cref{sec:conclusions}.

\section{Methods}
\label{sec:methods}
\subsection{Kinetic Inductance Detector}
\begin{table*}[ht!]
\centering
\caption{Optical filter configurations and radiation sources used for single photon counting at \qtylist{3.8;8.5;18.5;25}{\um}. The filter positions at \qtylist{3.8;8.5;18.5}{\um} correspond to the DR setup seen in \cref{fig:mkid}(c,d). The ADR setup at \qty{25}{\um} is described in Ref. \cite{baselmans2022ultra} (Fig. 2c). For detailed descriptions of the filters, see Appendix \ref{app:A}.}
\label{tab:filterconfigs}
\resizebox{\textwidth}{!}{%
\begin{tabular}{llllll}
\hline
\textbf{} &
   &
  \textbf{3.8 µm} &
  \textbf{8.5 µm} &
  \textbf{18.5 µm} &
  \textbf{25 µm (ADR)} \\ \hline
\multirow{6}{*}{\textbf{\begin{tabular}[c]{@{}l@{}}Filter \\ positions\end{tabular}}} &
  \textbf{Sample} &
  2$\times$BP$_{3.8}$+CaF$_{2}$ &
  2$\times$BP$_{8.5}$+CaF$_{2}$ &
  3$\times$BP$_{18.5}$ &
  BP$_{25}$+SP$_\mathrm{B}$ \\
 &
  \textbf{\qty{100}{mK}} &
  BP$_{3.8}$+$2\times$CaF$_{2}$ &
  BP$_{8.5}$+$2\times$CaF$_{2}$ &
  BP$_{18.5}$+ZnSe+ND$_1$ &
  BP$_{25}$+2$\times$SP$_\mathrm{A}$ \\
 &
  \textbf{ND \qty{100}{mK}} &
  ND$_3$ &
  ND$_3$ &
  ND$_3$ &
  - \\
 &
  \textbf{ND \qty{3}{K}} &
  ND$_2$ &
  ND$_2$ &
  - &
  - \\
 &
  \textbf{\qty{3}{K}} &
  BP$_{3.8}$+CaF$_{2}$ &
  BP$_{8.5}$+CaF$_{2}$ &
  BP$_{18.5}$+ZnSe &
  BP$_{25}$+SP$_\mathrm{A}$+LP \\
 &
  \textbf{\qty{30}{K}} &
  CaF$_{2}$ &
  CaF$_{2}$ &
  Blind flange &
  - \\
\textbf{Source} &
   &
  \begin{tabular}[c]{@{}l@{}}QTH lamp, \\ external\end{tabular} &
  \begin{tabular}[c]{@{}l@{}}Thermal background, \\ external\end{tabular} &
  \begin{tabular}[c]{@{}l@{}}Thermal radiator, \\ at \qty{3}{K}\end{tabular} &
  \begin{tabular}[c]{@{}l@{}}Thermal radiator, \\ at \qty{3}{K}\end{tabular} \\ \hline
\end{tabular}%
}
\end{table*}
MKIDs are superconducting detectors, where photons with energy greater than twice the gap energy \unit{\Delta} of the superconductor can break Cooper pairs and change the complex impedance of a superconducting film. By integrating the superconducting film in a microwave resonance circuit, the change in impedance can be read out as a change of the complex $S_{21}$ transmission at resonance \cite{day2003broadband}. 

The MKID array used in this work is identical to the one described in detail by \textcite{baselmans2022ultra}. Here we briefly reiterate the key design features with the single device architecture in \cref{fig:mkid}(a,b) and the full array configuration in \cref{fig:coincidence}a. The array consist of 27 MKIDs. Each MKID consists of a niobium-titanium-nitride (NbTiN) interdigitated capacitor (IDC) and an aluminium coplanar waveguide (CPW) with a highly inductive central line; the THz line. The detector is capacitively coupled to the readout line at the IDC side and shorted at the end of the THz line.
Radiation is coupled to the detector by a silicon (Si) lens that focuses the light onto the feed point of a leaky-slot antenna at the end of the THz line. Key is the suspension of the aluminium CPW on a $\sim\qty{100}{\nm}$ thick sillicon-nitride (SiN) membrane for re-trapping of $2\Delta$ phonons generated by quasiparticle recombination. The sensitivity and optical coupling have been optimized for the far-IR at \qty{200}{\um}. 

In this work we use the same detector at all wavelengths. The detector used has a \qty{0.773}{\um\tothe3} aluminum volume, \qty{4.60}{GHz} resonance frequency, $\qty{50}{\us}$ quasiparticle lifetime (in dark conditions) and \qty{2.3}{meV} energy resolution (from \cite{baselmans2022ultra}, Fig. 4d and Fig. B.1). This energy resolution promised the photon-counting ability in the mid-IR that has led to this work.

\subsection{Cryogenic setup and Microwave readout}
\label{sec:cryogenic setup}

\begin{figure*}[ht!]
    \includegraphics[width=\linewidth]{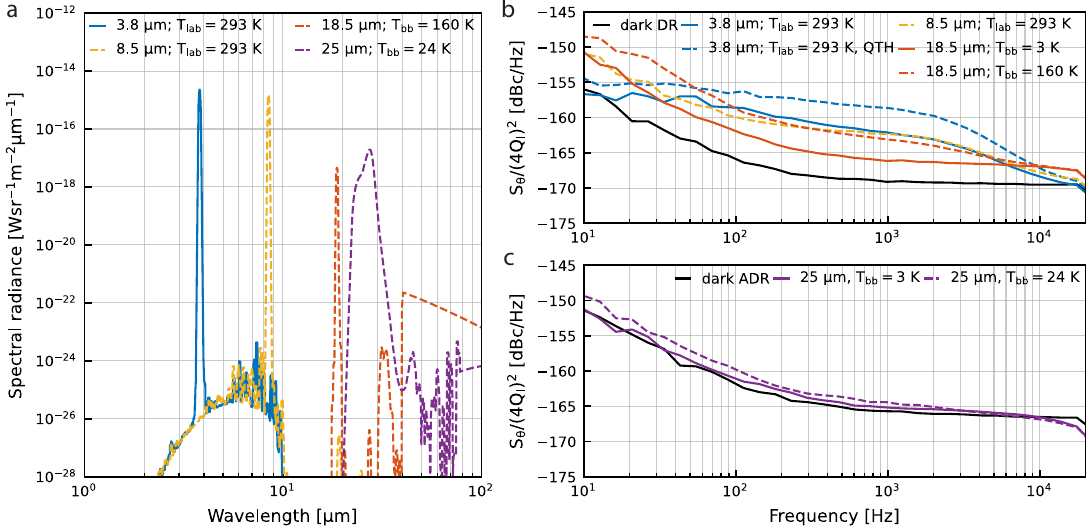}
    \caption{Characterization of the measurement setups. (a) Estimated spectral radiance at the detector and (b,c) measured fractional frequency noise at \qtylist[list-units=single]{3.8;8.5;18.5;25}{\um} (blue, yellow, orange, purple, respectively) and in the dark (black). The \qty{25}{\um} experiment was done in (c) the ADR, the others in the (b) DR. In all panels the solid lines represent the radiance and noise from just the background, while the dashed lines also include the radiation source.}
    \label{fig:setup characterization}
\end{figure*}

A well controlled measurement setup is required to demonstrate single-photon counting with such a sensitive MKID detector. To have clearly distinguishable single-photons, the setup should allow the detector to see about 
\numrange{100}{1000} photons per second; equivalent to \qtyrange[range-units=single]{1}{10}{aW} of power absorbed by the detector at \qty{18.5}{\um}. This requires an extremely light-tight setup as photons with an energy larger than the energy gap of aluminum (\qty{4}{mm} wavelength) can still break Cooper pairs in the detector. 

The detector array is located at the sample stage of a Bluefors dilution refrigerator (DR) at a typical temperature of \qty{100}{mK}. The cryostat is configured in one of two ways, see \cref{fig:mkid}(c,d). The differences between the configurations will be discussed in more detail in \cref{sec:optical setup}, but the common elements are discussed here. The detectors are shielded from higher temperature environments by the \qty{100}{mK} and subsequent \qtylist[list-units=single]{3;30}{K} shields. Shielding to external electric and magnetic fields are provided by a Cryophy and superconducting Niobium shield also at \qty{100}{mK}. The \qty{100}{mK} shield has a long snout containing a series of baffle rings for stray light control and the snout is also fully coated with a radiation absorbing layer of carbon-loaded epoxy with \unit{SiC} grains \cite{klaassen2001optical}. A set of optical filters can be mounted at the indicated positions. The vacuum window is made from \qty{6}{mm} thick CaF$_2$ and is not considered as an optical filter as it is at ambient temperature. Dedicated neutral density (ND) filter holders have been designed to prevent standing waves between the filters. The holders tilt the ND filters at a \qty{45}{\degree} angle, reflecting the radiation to an absorbing layer that covers the inside. The ND holders can be mounted to the \qty{100}{mK} and \qty{3}{K} stages.   
The \qty{25}{\um} measurements are done in the adiabatic dilution refrigerator (ADR) from Ref. \cite{baselmans2022ultra}. The configuration of the ADR is similar to the DR configuration at \qty{18.5}{\um} (\cref{fig:mkid}d) with a cryogenic radiator at \qty{3}{K}. 
To measure the dark count rates we perform a multichannel dark count measurement in the DR which is described in more detail in Appendix \ref{app:B}.

The microwave readout scheme used in this work is the same as the one described by Ref. \cite{kouwenhoven2024visible} with the \qtyrange[range-units=single, range-phrase=--]{2}{6}{GHz} LNA. The response of the detector is the change in the in-phase and quadrature components of a readout tone centered on the equilibrium (dark) detector resonance. For single-photon pulse measurements we generally work with the phase response $\theta$, sampled at \qty{1}{MHz}. The readout power $P_{\mathrm{read}}$ is optimized to just below the onset of bifurcation.

\subsection{Optical setup and characterization}
\label{sec:optical setup}

We require different photon sources and filtering solutions to define the exact wavelength bands of interest at the desired power. In \cref{fig:mkid}d we show the cryostat configuration at \qtylist[list-units=single]{18.5;25}{\um}. The detectors are illuminated by a cryogenic radiator mounted on the \qty{3}{K} stage that can be modeled as an ideal black body. The radiator can be heated by a resistor to temperatures up to \qty{180}{K} without causing to much heat load on the \qty{100}{mK} stage. Therefore, a different source is required at \qtylist[list-units=single]{3.8;8.5}{\um}. The source at these wavelengths is external to the cryostat and illuminates the detector through a series of optical filters, \cref{fig:mkid}c. This also means we have a thermal background present with a temperature of $T_\mathrm{lab}\approx\qty{293}{K}$ that peaks in radiance around \qty{10}{\um}. At \qty{3.8}{\um} we use a monochromator source: a \qty{100}{W} Quartz Tungsten Halogen (QTH) lamp in series with a monochromator (Oriel Cornerstone\texttrademark $260$) that spectrally filters the light using a tunable grating (model: 74080). The QTH lamp has no significant radiance at \qty{8.5}{\um} such that we only rely on the thermal background as our photon source.

We have summarized the filter configurations for the different wavelengths in \cref{tab:filterconfigs}. Common to the filter stacks are the band-pass (BP) filters that transmit the desired wavelength and reject the background. The BP filters are mounted at multiple temperature stages to reduce the effect of filter heating and re-radiation \cite{tucker2006thermal}. \unit{CaF_2} and \unit{ZnSe} filters are added for short pass filtering and Neutral Density (ND) filters are used to reduce the power at the detector to acceptable levels. It is likely that some of the absorptive filters become increasingly transparent for the far-IR and longer wavelengths. In Appendix \ref{app:A} we list the relevant characteristics of all the individual filters and provide relevant literature about their far-IR transmission.

We characterize the different measurement setups by computing the expected spectral radiance at the detector and by measuring the corresponding fractional frequency noise, see \cref{fig:setup characterization}.
The fractional frequency noise---hereafter denoted as \emph{noise}---is computed from the power spectral density $S_\theta$ and loaded quality factor $Q_l$ of the resonator as $S_\theta/(4Q_l)^2$. $S_\theta$ is computed from the resonator's phase response with Welch's method. $Q_l$ is obtained from a Lorentzian fit to the MKID's resonance dip. The fit incorporates both an asymmetric component according to  Ref. \cite{khalil2012analysis, probst2015efficient} as well as a non-linear component---as we operate near bifurcation---according to Ref. \cite{swenson2013operation}.

The expected spectral radiance $I$ as a function of wavelength $\lambda$ and black-body temperature $T_\mathrm{bb}$ is given by 
\begin{equation}
    I(\lambda; T_\mathrm{bb})=A\Omega B_\lambda(\lambda; T_\mathrm{bb})\Theta(\lambda),
\end{equation}
with $A=\pi D^2/4$ the surface area of the lens (diameter $D=\qty{1.550}{mm}$), $\Omega$ the solid angle  of the limiting aperture in the setup (\qty{2}{\degree} in the DR, \qty{9}{\degree} in the ADR) and $\Theta$ the total filter transmission. Lastly, $B_{\lambda}$ is the black-body radiance given by Planck's law:  
\begin{equation}
    B_\lambda(\lambda; T_{\mathrm{bb}}) = \frac{2hc^2}{\lambda^5}\frac{1}{e^{hc/(\lambda k_\mathrm{B} T_{\mathrm{bb}})-1}},
    \label{eq:planck}
\end{equation}
with $h$, $k_\mathrm{B}$ and $c$ physical constants. 
In \cref{fig:setup characterization}a we have plotted the spectral radiance for every setup and in \cref{fig:setup characterization}b we have plotted the corresponding noise levels. We observe the highest background noise at \qty{3.8}{\um} (QTH lamp off) and the lowest at \qtylist[list-units=single]{18.5;25}{\um} ($T_\mathrm{bb}=\qty{3}{K}$). We suspect far-IR radiation, transmitted through the filters, to cause the increased photon noise at \qty{3.8}{\um} and, to a lesser extent, at \qty{8.5}{\um}.

The \qtylist[list-units=single]{18.5;25}{\um} configurations have the lowest noise levels because a cryogenic source is used. However, there is a large difference in the required radiator temperatures: \qtylist[list-units=single]{160;24}{K}, respectively. The higher temperatures at \qty{18.5}{\um} are due to the narrower \qty{18.5}{\um} band pass filters, requiring higher radiator temperatures for the same in-band power. Consequently, there is much more background radiation ($P\propto T^4$) present in the \qty{18.5}{\um} setup which results in higher noise levels. Additionally, we expect the absorptive filters at \qty{18.5}{\um} to be more transparent for the far-IR than the reflective filters used at \qty{25}{\um}.

Additional characterization of the setup is done by measuring the photon count rate as a function of radiator temperature. No valid value for the optical efficiency can be extracted from this as it is nearly impossible to accurately model the optical efficiency of the lens-antenna structure of this detector in the mid-IR. Nonetheless, it is insightful to check whether the detector and setup behave as expected. We model the absorbed photon count rate $N_{\mathrm{ph}}$ as a function of $T_\mathrm{bb}$ by integrating the radiance from \cref{eq:radiance} and dividing by the photon energy $E_{\mathrm{ph}}$: 
\begin{equation}
    N_{\mathrm{ph}}(T_{\mathrm{bb}}) = \frac{\eta}{E_{\mathrm{ph}}} \int_{\mathrm{in\text-band}} I(\lambda;T_\mathrm{bb})\mathrm{d}\lambda.
    \label{eq:radiance}
\end{equation}
The \emph{in-band} integration range is defined as the full width (\qty{99}{\percent}) of the main transmission peak of the entire filter stack seen in \cref{fig:setup characterization}a. $\eta$ is the fitting parameter of our model and combines all the unknowns about the absorption mechanism of our detector in the mid-IR, including, but not limited to the optical efficiency. It is assumed that $\eta$ is constant within the integration range.

\subsection{Pulse detection and resolving power}
\label{sec:Pulse detection}
A single-photon response of the detector is recognized in the time domain data by a fast rising edge followed by a slow exponential decay. Pulse detection is done in three steps. First, we remove noise spikes from the data using a three-point moving median filter. Second, we increase the signal-to-noise by smoothing the data with a falling exponential window that has a lifetime that is iteratively matched to that of the detected pulses. Last, we set a threshold for pulse triggering. We quantify the pulse detection thresholds in this work by means of the standard deviation of the smoothed noise, $\sigma$. Practically, the pulses are detected using the \emph{scipy.signal.find\_peaks} \cite{scipy2026findpeaks} module in Python with both the \emph{height} and the \emph{prominence} attributes set to the threshold value. Generally, we set the pulse detection threshold right at the noise floor, \qty{3}{\sigma}.

After the pulses have been detected we want to find their pulse heights. The pulse-height distribution tells us about what type of pulses we have in our system and the spread of pulse heights determines the energy resolving power. The pulse heights in the presence of noise are most accurately estimated using an optimal filter constructed from the average pulse shape and average noise power spectral density (PSD) \cite{szymkowiak1993signal,irwin1995phonon}. The average pulse shape is constructed from a subset of all the detected pulses: the singe-photon pulses from photon absorbed directly in the aluminium central line. We will explain this further \cref{sec:resolving power}. A single-photon pulse is defined as a pulse with no consecutive detections within a \qty{400}{\us} time window. The average noise PSD is constructed from a set of \qty{400}{\us} noise segments cut from the time stream in between the pulses.

\begin{figure*}[ht!]
    \centering
    \includegraphics[width=\linewidth]{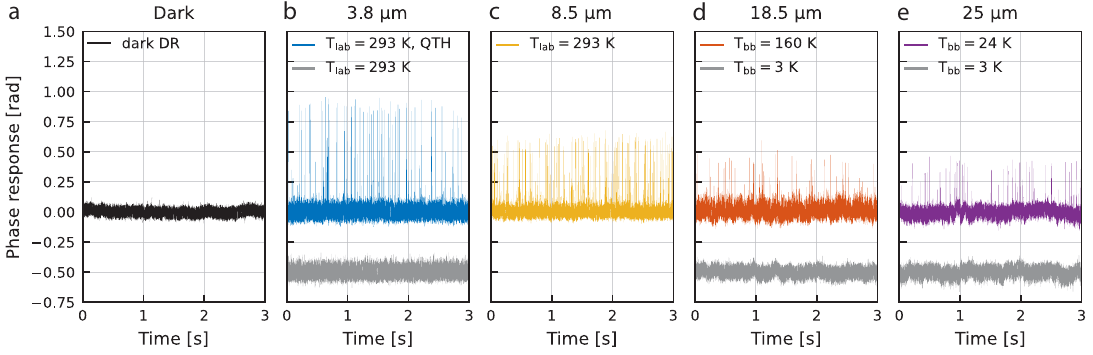}
    \caption{Single-photon detection across the mid-IR. Shown are phase response time streams in the dark and at \qtylist[list-units=single]{3.8;8.5;18.5;25}{\um} in panels \emph{a-e}, respectively. At \qtylist[list-units=single]{3.8;18.5;25}{\um} we show the absence of pulses when the radiation source is turned \emph{off} (gray line, offset by \qty{-0.5}{rad} for clarity). At \qty{8.5}{\um} the radiation source is the every present thermal background of the lab. The data is smoothed with an exponential window with a lifetime of \qty{50}{\us}. We rescale the response at a wavelength $x$ by a factor $Q_\mathrm{l}^x/Q_\mathrm{l}^\mathrm{dark}$.} 
    \label{fig:photon pulses}
\end{figure*}

The energy resolving power $R=E/\delta E$ shows with what energy resolution $\delta E$ the detector resolve the photon energy $E$. The response of the detector scales linearly with photon energy for the wavelengths measured in this work such that we can write $R=\bar{H}/\delta H$ with $\bar{H}$ average pulse height and $\delta H$ the full-width-at-half-maximum of the pulse-height distribution. We estimate $\delta H$ with a Kernel Density Estimate (KDE). We determine the resolving power for the same selection of pulses as used for the optimal filter. To characterize the detector and measurement setup we want to identify what the limiting contribution is to the resolving power. We define three possible contributions to the resolving power. First, the spectral (im)purity of the source and filter stack. This contribution is very small for the results of this work and neglected. The second contribution is the signal-to-noise ratio of the detector and readout chain, $R_\mathrm{SN}=\bar{H}/\delta H_\mathrm{noise}$ with $\delta H_\mathrm{noise}$ the width of the optimally filtered noise. The last contribution is generally indicated as the \emph{intrinsic} energy resolution and is obtained from the others $1/R_\mathrm{i}^2 = 1/R^2 - 1/R_\mathrm{SN}^2$. So, $R_\mathrm{i}$ includes all intrinsic effects that limit the resolving power, except for the detector noise. The ultimate limit to the resolving power is given by the statistical variation in the number of quasiparticles generated per photon; the Fano limit $R_\mathrm{Fano}$. Device-specific limitations to the resolving power are captured by $R_\mathrm{phonon}$; a measure of how much the photon's energy is kept inside the detector and not lost by phonons that escape during the predetection stage\cite{devisser2021phonon}. $R_\mathrm{phonon}$ is given by
\begin{equation}
    R_\mathrm{phonon} =\frac{1}{2\sqrt{2\ln 2}}\sqrt{\frac{\eta^\mathrm{max}_\mathrm{pb}E}{\Delta (F+J)}}
    \label{eq:Fano}
\end{equation}
with $E$ the photon energy and $\eta^\mathrm{max}_\mathrm{pb}=0.59$ the pair-breaking efficiency\cite{kozorezov2000quasiparticle}. $F \approx 0.2$ is the Fano factor for superconductors \cite{rando1992properties, kurakado1982possibility}. Note that $R_\mathrm{Fano}=R_\mathrm{phonon}(J=0)$. Membrane devices thus have the potential to outperform devices on solid substrates by about a factor \num{2.4} in resolving power \cite{devisser2021phonon}, but whether this also holds in the mid-IR is unkown.

\section{Results}
\label{sec:results}
\subsection{Single-photon detection across the mid-IR}
In \cref{fig:photon pulses} the main result of this work is shown: single-photon pulses across the mid-IR. Visible are five \qty{1}{s} time streams, starting on the left with the dark configuration and continuing to the right with the configurations at \qtylist[list-units=single]{3.8;8.5;18.5;25}{\um}, respectively. We have smoothed the data with a falling exponential window with a lifetime that is approximately equal to that of the pulses. This differs between \qtyrange{50}{70}{\us} between the configurations due to the different background loading. We also compensate for the difference in $Q_l$ by rescaling the response at a wavelength $x$ by a factor $Q_\mathrm{l}^x/Q_\mathrm{l}^\mathrm{dark}$. At \qty{3.8}{\um} we clearly see the pulses disappear when the QTH lamp is turned off (gray line, offset by \qty{-0.5}{rad} for visibility). Similarly, at \qtylist[list-units=single]{18.5;25}{\um} the pulses disappear when the radiator is turned off. At \qty{8.5}{\um} we just see the photons that originate form the thermal background. In \cref{fig:resolving power}b we have plotted the mean pulse shape of the pulses at \qty{3.8}{\um} both on linear and semi-logarithmic scale (inset). The quasi-particle lifetime $\tau_\mathrm{qp}$ of the pulse is obtained from fitting $y=a\mathrm{e}^{-x/\tau_\mathrm{qp}}$ with $a,b$ and $\tau_\mathrm{qp}$ the fitting parameters. We find $\tau_\mathrm{qp}=\qty{55\pm1}{\us}$.

\subsection{Energy resolving power}
\label{sec:resolving power}
\begin{figure*}[ht!]
    \centering
    \includegraphics[width=\linewidth]{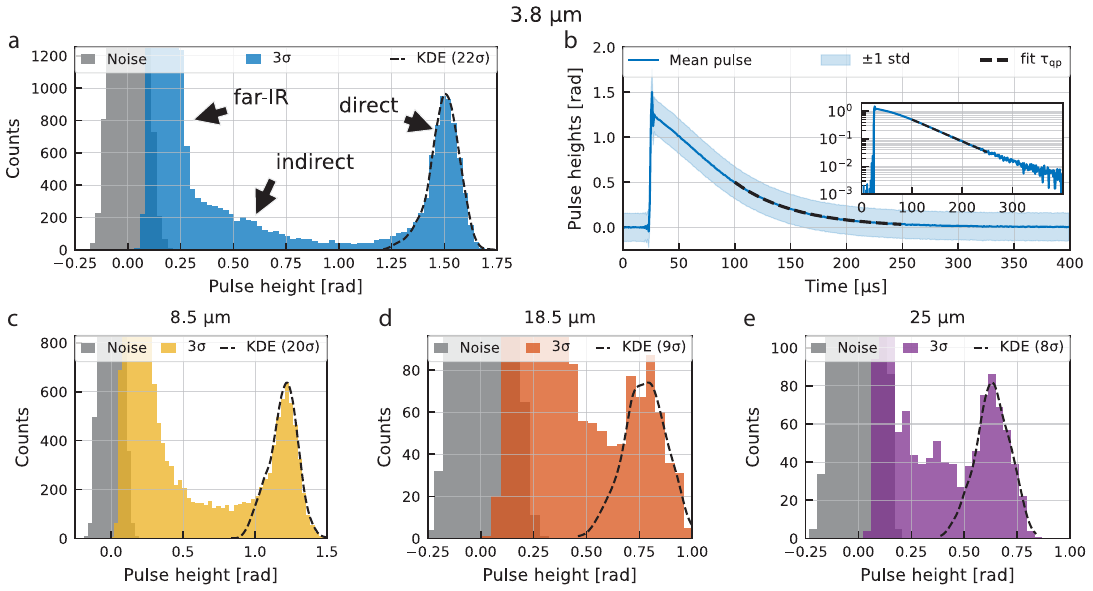}
     \caption{Energy resolving power in the mid-IR. Optimally filtered noise (gray shaded) and pulse-height distributions at (a, blue shaded) \qty{3.8}{\um} (c, yellow shaded) \qty{8.5}{\um} (d, orange shaded) \qty{18.5}{\um} and (e, purple shaded) \qty{25}{\um}. We have annotated three different pulse contributions in panel \emph{a} which are explained in main text. The pulses are detected with a \qty{3}{\sigma} threshold in the smoothed data and then cut from the raw data before being optimally filtered. The optimal filter is constructed from the pulses in the primary distribution which is identified by the KDE (black, dashed line) and obtained by increasing the detection threshold to the level in the legend. The resolving power is then determined for the same selection. We rescaled the pulse heights at a wavelength $x$ by a factor $Q_\mathrm{l}^x/Q_\mathrm{l}^\mathrm{dark}$. (b) The average pulse shape of the directly absorbed \qty{3.8}{\um} photons (indicated by the KDE in panel \emph{a}), both on a linear and semi-logarithmic scale (inset). The shaded region indicates the $\pm1$ standard deviation from the average pulse shape. An exponential decay fitted to the tail of the pulse (black, dashed line) gives a $\tau_\mathrm{qp}=\qty{55\pm1}{\us}$.}
    \label{fig:resolving power}
\end{figure*}

We investigate the MKID pulse detection performance in the mid-IR by measuring the energy resolving power and its contributions. We plot the optimally filtered noise and pulse-height distributions at \qtylist[list-units=single]{3.8;8.5;18,5;25}{\um} in panels \emph{a}, \emph{c}, \emph{d} and \emph{e} of \cref{fig:resolving power}, respectively. We can distinguish three different pulse contributions: \emph{direct}, \emph{indirect} and \emph{far-IR} pulses. These are most clearly observed at \qty{3.8}{\um} (\cref{fig:resolving power}a, annotations). The \emph{direct} pulses are from single-photons absorbed in the aluminium central line and make up the primary distribution above \qty{1}{rad}. The \emph{indirect} pulses are from single-photons absorbed in the aluminium groundplane. These can lead to a pulse response in two ways: (1) the photon energy generates quasiparticles in the ground plane, creating a pulse with much smaller responsivity than a pulse from a photon absorbed in the central line; and (2) the photon energy is converted to phonons, which partially reach the central line and again lead to a pulse response, but with a smaller pulse height. The indirect pulses were also observed by Ref. \cite{devisser2021phonon} (Appendix B). The last pulse contribution is from the \emph{far-IR} background that leads to a large number of detections just above the noise floor. When comparing the far-IR pulse contributions at all wavelengths one can see that it is smallest at \qty{25}{\um} (\cref{fig:resolving power}e). This is as expected from \cref{fig:setup characterization}. 
The resolving power of the detector is determined from the direct pulses alone, as is the optimal filter. The direct pulses are selected by increasing the pulse detection threshold such that we remain only with the primary pulse-height distribution. The resolving power then is determined with a KDE. The KDEs for all the wavelengths are plotted in \cref{fig:resolving power} with their corresponding pulse detection threshold indicated in the legends. Note that the detection threshold is set in the smoothed data, before optimal filtering; the optimal filtering transforms the hard cut in smoothed pulse heights into a gradual transition seen from the KDEs. This method enables us to determine the width---and thus resolving power---of the direct pulse distribution even when it is obscured by the indirect pulses, as is the case at \qty{18.5}{\um} (\cref{fig:resolving power}d).

\begin{figure*}[ht!]
    \centering
    \includegraphics[width=\linewidth]{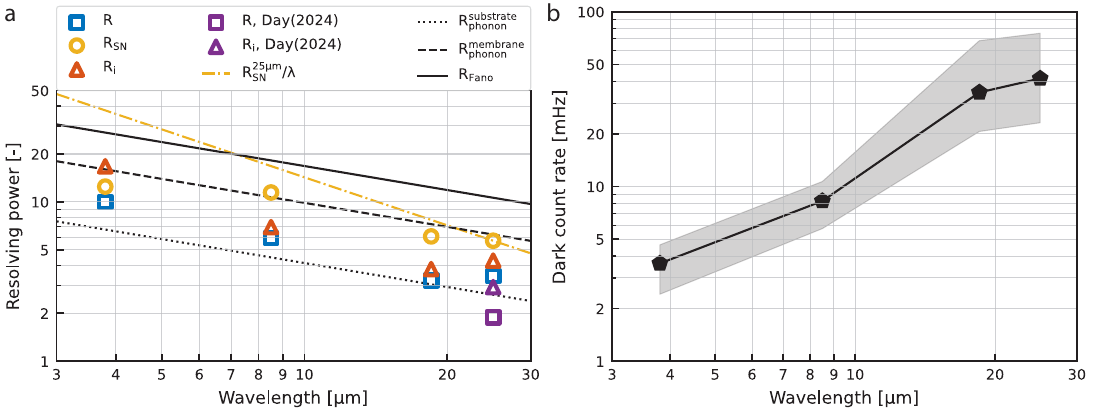}
    \caption{Energy resolving powers and dark count rates at \qtylist[list-units=single]{3.8;8.5;18.5;25}{\um}. (a) Plotted are the energy resolving power $R$ and the signal-to-noise $R_\mathrm{SN}$ and intrinsic $R_\mathrm{i}$ contributions (blue, yellow and orange markers, respectively) that were obtained at \qtylist[list-units=single]{3.8;8.5;18.5;25}{\um}. We plot the Fano limit from \cref{eq:Fano} (black, solid line, $J=0$) and the phonon-loss limited resolving powers for solid-substrate ($J=3.1$; black, dotted line) and membrane devices ($J=0.38$; black, dashed line). We extrapolate the $R_\mathrm{SN}$ at \qty{25}{\um} to the other wavelengths by scaling with $\lambda$ (yellow, dash-dotted line). We also plot $R$ and $R_\mathrm{i}$ obtained by Ref. \cite{day202425} at \qty{25}{\um} (purple markers). (b) Dark count rates obtained from a \qty{10000}{s} measurement, see also Appendix \ref{app:B}. The dark count rate are calculated per wavelength using the pulse-height distributions in \cref{fig:resolving power} and a \qty{99.73}{\percent} ($\mu\pm 4\sigma$ or $\bar{H}\pm 1.7\delta H$) confidence interval. The shaded area gives the \qty{95.45}{\percent} ($\mu\pm 3\sigma$) to \qty{99.99}{\percent} ($\mu\pm 5\sigma$) confidence range.}
    \label{fig:fano}
\end{figure*}

The results of $R$, $R_\mathrm{SN}$ and $R_\mathrm{i}$ at all wavelengths measured in this work are plotted in \cref{fig:fano}a. We also plot the phonon-limited resolving power $R_\mathrm{phonon}$ for a substrate ($J=3.1$, dashed line) and membrane ($J=0.38$, solid line) device, see also \cref{sec:Pulse detection}. The values for $J$ are obtained in the visible/near-IR regime from Ref. \cite{devisser2021phonon} and extrapolated to the mid-IR (using $T_c=\qty{1.54}{K}$). Lastly, we plot the results from Ref. \cite{day202425} at \qty{25}{\um}. At \qty{3.8}{\um} we find that $R_\mathrm{i}$ matches $R_\mathrm{phonon}$ for membrane devices. This indicates that phonon-trapping affects the resolving power at our mid-IR wavelengths in the same way as was found for higher energy photons, as expected. 
We do not reach phonon-loss limited performance at other wavelengths due to the indirect photon pulses broadening the direct pulse-height distribution. The large contribution of indirect detections is a consequence of the unoptimized optical coupling. While the detector intrinsically performs at its best at \qty{3.8}{\um}, the measurement setup seems to limit the resolving power. This is illustrated best when we extrapolate the $R_\mathrm{SN}$ found at \qty{25}{\um} to the lower wavelengths by rescaling with the photon energy, see the yellow, dash-dotted line in \cref{fig:fano}a. The line shows that we have the highest noise in the setup at \qty{3.8}{\um} which was also observed in \cref{sec:optical setup} and attributed to the far-IR background.

\subsection{Dark count rate}
We have measured the dark count rate for the detector, see \cref{fig:fano}b. The dark count measurement and analysis are described in Appendix \ref{app:B}. We use the energy resolving power to discern a dark count rate per wavelength and find values of \qtylist[list-units=single]{4;8;35;48}{mHz} at the wavelengths of \qtylist[list-units=single]{3.8;8.5;18.5;25}{\um}, respectively. These values are determined with the resolving power of the detector at every wavelength and by requiring a \qty{99.73}{\percent} (i.e. $\mu\pm 4\sigma$ or $\bar{H}\pm 1.7\delta H$) confidence to detect all the direct single-photon pulses. A higher resolving power will lower the dark count rate further: a phonon-loss limited resolving power for a membrane device at \qty{18.5}{\um} would lead to dark count rates of around \qty{5}{mHz} instead of \qty{20}{mHz} in our current setup.

\begin{figure}[hb!]
    \centering
    \includegraphics[width=1\linewidth]{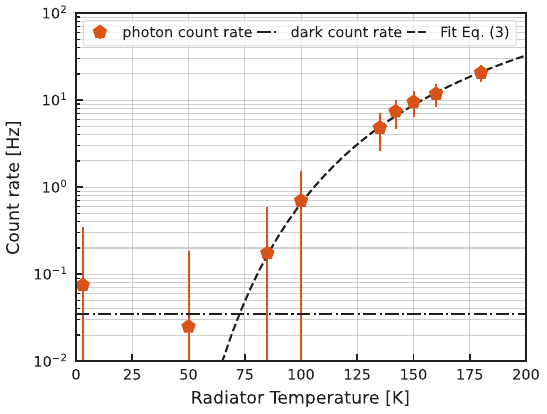}
    \caption{Photon count rates $N_\mathrm{ph}$ versus radiator temperature $T_{\mathrm{bb}}$ at \qty{18.5}{\um}. The photon count rate is obtained from \qty{40}{s} measurements. A $9\sigma$ detection threshold was applied to only detect direct photon pulses, see also \cref{fig:resolving power}d. No photons were detected at \qty{30}{K}. The error bars are given by the square root of the mean. We fit \cref{eq:radiance} to the data at $T_{\mathrm{bb}}\geq\qty{85}{K}$ and find $\eta = (7.8 \pm 0.2) \times 10^{-2}$. We plot the dark count rate from \cref{fig:fano}b at \qty{18.5}{\um} to show consistency.}
    \label{fig:Nph}
\end{figure}

Lastly, we validate the behavior of the detector and measurement setup at \qty{18.5}{\um}, see \cref{fig:Nph}. We measure the number of direct photon pulses versus radiator temperature using the \qty{9}{\sigma} detection threshold also used in \cref{fig:resolving power}d.  The model from \cref{eq:radiance} fits very well to the measured photon rates $N_\mathrm{ph}$ at $T_\mathrm{bb}\geq\qty{85}{K}$ using a single fitting parameter $\eta = (7.8 \pm 0.2) \times 10^{-2}$. We have included the transmission of the Si lens into this calculation which is around \num{0.2} at \qty{18.5}{\um}, see \cref{fig:filters}b (from Ref. \cite{wollack2020infrared}).

\section{Conclusions}
\label{sec:conclusions}
We have demonstrated single-photon counting at \qtylist[list-units=single]{3.8;8.5;18.5;25}{\um} using an MKID with aluminium on a thin membrane. At every wavelength a dedicated setup was designed, utilizing different radiation sources and filter configurations to cover the broad range of wavelengths. We measured energy resolving powers $R$ of \numlist{9.9;5.9;3.2;3.3}, respectively. With these resolving powers we obtained dark count rates of \qtylist[list-units=single]{4;8;35;48}{mHz}, respectively, from a \qty{10000}{s} dark measurement.

Higher resolving powers are desirable to reduce the dark count rates of the detector. We reach phonon-loss limited performance at \qty{3.8}{\um} and we believe this performance can be matched across the mid-IR, outperforming similar solid substrate devices by about a factor \num{2.4} \cite{devisser2021phonon}. We have investigated the current limitations to the resolving powers measured in this work.
At \qty{3.8}{\um} we are limited by an excess far-IR background in the setup. This is not surprising as we are exposing a single-photon sensitive detector to a \qty{293}{K} thermal environment that requires---even in-band---more than \qty{50}{dB} attenuation. The background radiation can be decreased by adding metal-mesh short pass filters or by using a cryogenic source like mid-IR Quantum Cascade Lasers \cite{yao2012mid}. At \qtylist[list-units=single]{8.5;18.5;25}{\um} we are limited by pulses from photons absorbed in the aluminium ground plane. This can be eliminated in future devices by using a lens-absorber coupling structure that is optimized for the mid-IR \cite{dabironezare2025lens,day202425}. This will likely require an optimized absorber per spectral channel as the mid-IR spans a factor \num{4} in wavelength. Besides, at \qty{18.5}{\um} the far-IR background should also be reduced by using broader band pass filters such that lower radiator temperatures can be used while having the same in-band power.
To conclude, phonon-loss limited performance with membrane MKIDs is within reach across the mid-IR. This will enable single-photon counting with extremely low dark count rates across the mid-IR.

\textbf{Data and code availability} The raw data and code are made available on Zenodo to reproduce all the results in this work: \url{https://doi.org/10.5281/zenodo.18714817} and  \url{https://doi.org/10.5281/zenodo.18714481}, respectively.

\begin{acknowledgments}
    This work is financially supported by the Netherlands Organisation for Scientific Research NWO (Vidi 213.149 and OCENW.M.23.028). We would like to thank Nick de Keijzer for designing adjustments to the setup for this experiments, and the SRON workshop for producing those.
\end{acknowledgments}

\appendix

\section{Optical filters}
\label{app:A}
\begin{figure}[ht!]
    \centering
    \includegraphics[width=\linewidth]{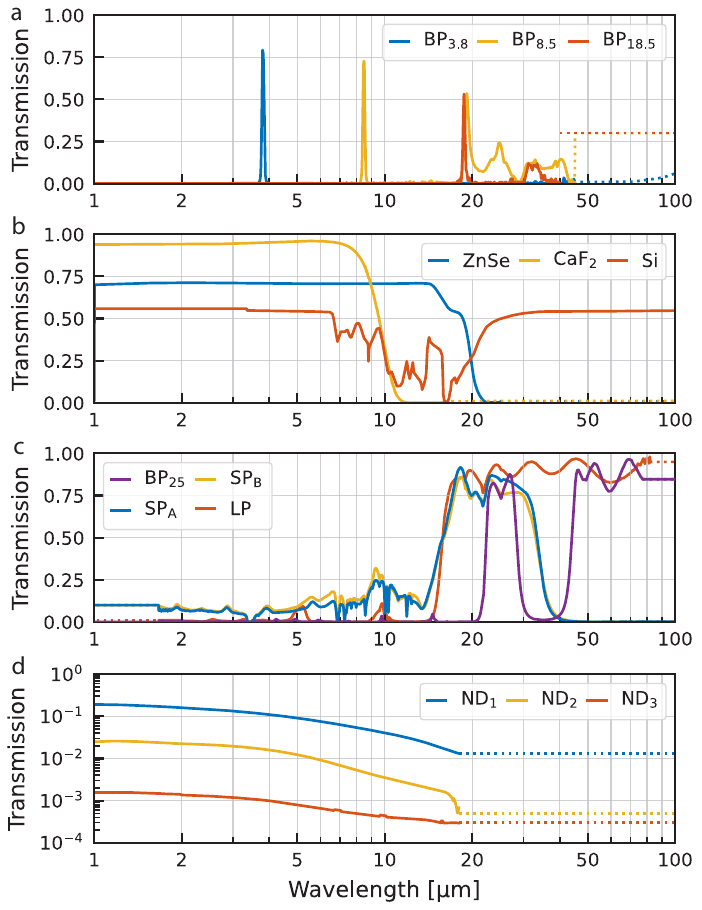}
    \caption{Transmission spectra of the filters used in the various measurement setups. The dotted parts of the spectra indicate the wavelength regions where the filter has not been characterized. Outside of the characterized region we estimate a worst-case transmission based on literature.}
    \label{fig:filters}
\end{figure}

Here we provide more detailed information on the specific filters used in the various measurement setups. In \cref{fig:filters} the transmission curves of all the individual filters are shown and in \cref{tab:filters} further specifications are listed. Based on literature we make an upper bound estimation for the optical transmission of the filters in the wavelength region not characterized by the manufacturer, up to \qty{100}{\um}. The consulted literature is indicated in the row \emph{Estimated FIR transmission} in \cref{tab:filters}.

\section{Dark counts}
\label{app:B}
\begin{figure}[htbp]
    \centering
    \includegraphics[width=\linewidth]{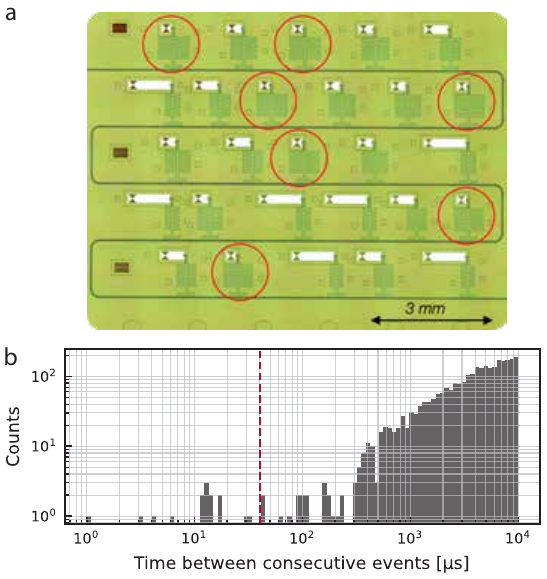}
    \caption{Coincidence triggering. (a) Spatial configuration of the \num{7} detectors used for the coincident triggering. One detector produced unusable data. (b) Time difference between consecutive events of all detector that were detected during a \qty{10000}{s} measurement. A threshold of \qty{40}{\us} yields only 8 coincident events leading to a total of 25 detections across all detectors ($< \qty{0.1}{\percent}$ of all events)}
    \label{fig:coincidence}
\end{figure}

We want to minimize the effect of cosmic rays to obtain an accurate estimate of the detector dark count rate. The energy resolving power of the MKIDs enables rejection of the primary, high energetic events based on their pulse height. However, the deposited energy of the primary event can spread across the array during its down conversion process, leading to secondary detections in other detectors that could be indistinguishable from single photon pulses. Therefore, we do a simultaneous measurement of \num{7} detectors and reject all coincident events; events that occur within \qty{40}{\us} across multiple detectors, similar to Ref. \cite{day202425}.
The multichannel measurement is done in the DR setup in the dark configuration, see also \cref{sec:methods}. At a sampling rate of \qty{1}{MHz} the readout was limited to a set of \num{7} detectors at the highest resonance frequencies (\qtyrange{4.39}{4.60}{GHz}) which includes the detector analyzed in the main text. The detectors were measured simultaneously for a duration of \qty{10000}{s}. The detectors are spatially distributed over the whole array, with some being nearest neighbors and others spaced further apart, see \cref{fig:coincidence}a. One detector produced unusable data, but its individual location is unknown due to possible swapping of the resonances. The remaining \num{6} detectors are analyzed identically and in the same way as the detector in the main text, see \cref{sec:Pulse detection}. We set a pulse detection threshold of \qty{5}{\sigma} to include all direct detections as well as many indirect detections (from cosmic rays), see \cref{fig:resolving power}.
We find \num{25} coincident detections that occurred in \num{8} separate events. This is a negligible amount of the total number of detections and hardly influences our dark count rates. The highly suppressed effect of cosmic rays was also observed by Ref. \cite{karatsu2015development} for a similar device with an aluminium groundplane on a membrane. The suppressed effect of cosmic rays probably limits our dark count rates because it rarely leads to a coincident event.
For the detector analyzed in the main text we remove \num{8} detections out of \num{12843}. We optimally filter the dark counts using the same pulse template as used for the resolving power but with the noise from the dark measurement. The dark count rates are calculated per wavelength by setting a confidence interval for the pulse height based distributions of directly absorbed photons in \cref{fig:resolving power}. The confidence interval gives the probability of including all the single-photons that were detected at that wavelength. The confidence interval is calculated on the mean and FWHM of the pulse height distributions shown in \cref{fig:resolving power}. A \qty{99.73}{\percent} (i.e. $\mu\pm 4\sigma$) confidence interval equals $\bar{H}\pm 1.7\delta H$. The dark count rates are given in \cref{fig:fano}b.

\begin{figure}[htbp]
    \centering
    \includegraphics[width=\linewidth]{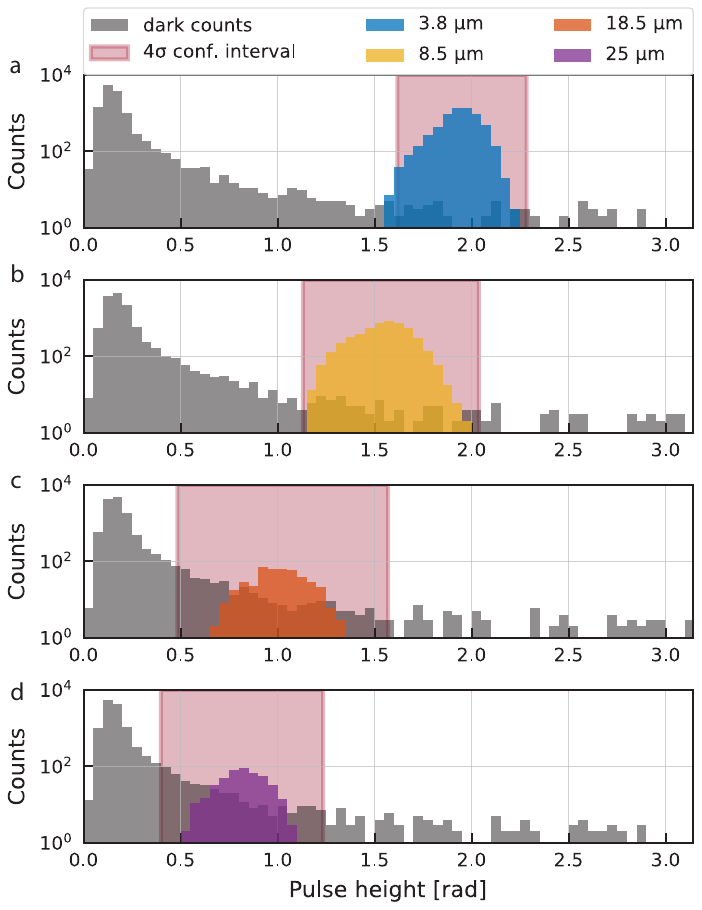}
    \caption{Dark count analysis at \qtylist[list-units=single]{3.8;8.5;18.5;25}{\um} (panels \emph{a-d}, respectively). The dark counts (gray shaded) were obtained during a \qty{10000}{s} dark measurement and optimally filtered using the noise from the dark measurement and the pulse templates per wavelength as also used for \cref{fig:resolving power}. The dark count rate per wavelength is determined from the overlap of the dark counts with the direct single-photon pulse distributions (blue, yellow, orange and purple shaded in top to bottom panels, respectively; from \cref{fig:resolving power} and scaled with $Q_\mathrm{l}^\mathrm{dark}$). The filled, red area represents the \qty{99.73}{\percent} ($\mu\pm 4\sigma$ or $\bar{H}\pm 1.7\delta H$) confidence interval of selecting all the direct single-photon pulses.}
    \label{fig:dark counts}
\end{figure}
% \clearpage
\begin{table*}[t]
\centering
\caption{Specifications of all the filters used in this work.}
\label{tab:filters}
\resizebox{\textwidth}{!}{
\begin{tabular}{@{}lllllllllll@{}}
\toprule
 &
  \textbf{BP$_{3.8}$} &
  \textbf{BP$_{8.5}$} &
  \textbf{BP$_{18.5}$} &
  \textbf{BP$_{25}$} &
  \textbf{CaF$_{2}$} &
  \textbf{ZnSe} &
  \textbf{ND$_{(1/2/3)}$} &
  \textbf{SP$_\mathrm{(A/B)}$} &
  \textbf{LP} &
  \textbf{Si} \\ \midrule
Manufacturer &
  NOC$^a$ &
  NOC$^a$ &
  NOC$^a$ &
  QMC$^b$ &
  EKSMA$^c$ &
  Thorlabs$^d$ &
  Thorlabs$^d$ &
  QMC$^b$ &
  QMC$^b$ &
  Veldlaser$^e$ \\
Part number &
  NC-2117 &
  NC-2118 &
  NC-2444-1 &
  FP3223 &
  530-6251 &
  WG70530 &
  NDIR(10/20/30)B &
  FP3238/FP3251 &
  FP3253 &
  - \\
Type &
  NBP$^f$ &
  NBP$^f$ &
  NBP$^f$ &
  BP$^g$ &
  SWP$^h$ &
  SWP$^h$ &
  ND$^i$ &
  SWP$^h$ &
  LWP$^j$ &
  Lens \\
cwl$^k$/fwhm$^l$  (\um) &
  3.8/76 &
  8.5/170 &
  18.5/370 &
  25/6300 &
  - &
  - &
  - &
  - &
  - &
  - \\
Cut-on (\um) &
  - &
  - &
  - &
  36 &
  - &
  - &
  - &
  - &
  14 &
  - \\
Cut-off  (\um) &
  - &
  - &
  - &
   &
  10 &
  22 &
  - &
  50 &
  - &
  - \\
Thickness (mm) &
  1.0 &
  1.0 &
  1.0 &
  $<$1 &
  1.0 &
  3.0 &
  1.0 &
  $<$1 &
  $<$1 &
  $^m$ \\
Diameter (mm) &
  25.4 &
  25.4 &
  25.4 &
  25.4 &
  25.4 &
  25.4 &
  25.4 &
  25.4 &
  25.4 &
  1.550 $^n$ \\
Substrate &
  SiO$_2$ &
  Ge &
  Ge &
  $^o$ &
  CaF$_2$ &
  ZnSe &
  ZnSe$^p$ &
  $^o$ &
  $^o$ &
  Si \\
Char. temp. (K) &
  77$^q$ &
  77$^q$ &
  77$^q$ &
  Room &
  Room &
  Room &
  Room &
  Room &
  Room &
  10 \\
Char. band (\um) &
  $0.8-45$ &
  $0.8-45$ &
  $0.8-40$ &
  $1.7-77$ &
  $0.2-16$ &
  $1-25$ &
  $0.2-18$ &
  $1.7-77$ &
  $1.7-82$ &
  $3-100$ \\
\begin{tabular}[c]{@{}l@{}}Estimated THz\\ transmission\end{tabular} &
  \begin{tabular}[c]{@{}l@{}}$0.7$ for \\ $\lambda>\qty{100}{\um}$ \cite{tydex2026thzmaterials}\end{tabular} &
  \begin{tabular}[c]{@{}l@{}}$0.3$ for \\ $\lambda>\qty{45}{\um}$ \cite{Kaplunov2019Ger}\end{tabular} &
  \begin{tabular}[c]{@{}l@{}}$0.3$ for\\ $\lambda>\qty{40}{\um}$ \cite{Kaplunov2019Ger}\end{tabular} &
  extrapolated$^r$ &
  \begin{tabular}[c]{@{}l@{}}$0.5$ for \\ $\lambda>\qty{200}{\um}$ \cite{Kaplunov2021CaF2}\end{tabular} &
  $0.01$ \cite{deneuville1991optical} &
  extrapolated$^r$ &
  extrapolated$^r$ &
  extrapolated$^r$ &
  $0.55$ \cite{wollack2020infrared} \\ \bottomrule
\end{tabular}
}
\justifying
\footnotesize{$^a$: Northumbria Optical Coatings ltd; $^b$: QMC Instruments ltd; $^c$:  EKSMA Optics; $^d$: Thorlabs, Inc.; $^e$: Veld Laser Innovations BV; $^f$: narrow band pass; $^g$: band pass; $^h$: short wave pass; $^i$: neutral density; $^j$: long wave pass; $^k$: central wavelength of the pass band; $^l$: full width at half maximum of the pass band; $^m$: lenses are elliptically shaped; $^n$: diameter of single lens; $^o$: the filter is composed of a stack of thin Copper meshes; $^p$: The ZnSe substrate is convered with a thin Nickel film; $^q$: characterized in-band at \qty{77}{K}, out-of-band at room temperature; $^r$: the last value of the characterized band is taken as the transmission at longer wavelengths;}
\end{table*}

\bibliography{bibliography}

\end{document}